\documentclass[a4paper]{article}
\usepackage{amsfonts}
\usepackage{amssymb}
\usepackage[margin=2cm]{geometry}
\usepackage[dvips]{graphicx}

\usepackage[T1]{fontenc}
\usepackage[latin2]{inputenc}

\usepackage{amsmath}
\usepackage{bbm}

\begin{document}


\title{A lattice gas model of a single Cooper pair box}

\author{Robert Alicki \\ 
  {\small
Institute of Theoretical Physics and Astrophysics, University
of Gda\'nsk,  Wita Stwosza 57, PL 80-952 Gda\'nsk, Poland}\\
}

\date{\today}
\maketitle

\begin{abstract}
There exists a large number of experimental and theoretical results supporting the picture of "macroscopic qubits" implemented by nanoscopic Josephson junctions. On the other hand the standard model of such systems given in terms of a single degree of freedom suggests their semiclassical behavior due to a localization mechanism caused by a strong coupling to an environment. Indeed, such a mechanism is observed in an atomic Bose-Einstein condensate (BEC) placed in a double-well potential - a system mathematically equivalent to a Josephson junction. In this note it is shown, on the example of a Cooper pair box,  that  replacing the BEC-type model for Cooper pairs by a lattice gas model one can reduce the environmental effects of  "dequantization" and can explain the experimental data in particular the existing huge differences between the measured values of relaxation times.
\end{abstract}
\section{Introduction}
In the last decade remarkable experiments were performed involving measurements and manipulations of states for  a single or several nanoscopic Josephson junctions which were consistently interpreted in terms of two level quantum systems \cite{W,Nak,Bla,Leh,Ans}. The main assumption in the theoretical analysis is that such a many-body mesoscopic system can be effectively treated as a quantum system of a single degree of freedom typically described by a large spin or nonlinear oscillator model. A nonlinear Hamiltonian yields the structure of two lowest energy levels which at the low enough temperatures can be separated from the others to form an effective \emph{macroscopic qubit}. The main problem with such models is a presence of a typically strong and collective
coupling to an environment. Namely, it is  expected that the observed states should be  rather well-localized semiclassical ones, which seem to be the only relatively stable with respect to  external noise. This mechanism in briefly discussed in the next Section. However, the semiclassical states for the model of Cooper pair box (CPB) are characterized  by large charge fluctuations which are not observed in the experiments. Therefore, either environmental decoherence producing semiclassical states does not work for Josephson qubits at the typical time scale of the experiments or the standard single degree of freedom model is not correct. The first alternative seems to be unlikely because the semiclassical character of observed states is confirmed in the recent experiments on atomic Bose-Einstein condensate (BEC) in a double-well potential \cite{Es}. Although this is a physically different system its mathematical description is the same as for the standard model of a CPB. In the Section 3 the second alternative is discussed, a lattice gas model of CPB. In this model the coupling to an environment has individual and local character and the effective size of the system is much smaller than in the standard model. Therefore, the picture of an \emph{ effective qubit}   
can be correct and some quantum features can be present on the relevant time scale. The price we pay is that the qubit states are not as well defined as in the standard model and strong leakage processes to other states in the effective Hilbert space are present.

\section{Collective single degree of freedom model}
The most studied, both experimentally and theoretically, examples of mesoscopic system which should support a qubit are "superconducting qubits". For simplicity, only a  CPB called "charge qubit" is discussed here. It is a circuit consisting of a small superconducting island connected  via Josephson junction to a large superconducting reservoir. Coulomb repulsion between Cooper pairs in a small electrode become important and must be taken into account in the Hamiltonian. The simple Josephson Hamiltonian reads \cite{W}
\begin{equation}
\hat{H} = 4E_C\sum_n( n- n_0 - n_g)^2 |n\rangle\langle n|  - \frac{E_J}{2}\sum_n ( |n+1\rangle\langle n| +|n\rangle\langle n+1|)
\label{cpbham}
\end{equation}
where $|n\rangle$ describe the state with $n$ Cooper pairs on the island, $E_C$ determines the magnitude of the Coulomb repulsion, $E_J$ governs the tunneling process,  $n_0 >> 1$ is a number of Cooper pairs on the island at the neutral reference state and the additional $n_g$ is a fine tuning of the external control. To establish a typical value of $n_0$ is a subtle problem, some authors assume that all Cooper pairs should be taken into account what yields $n_0\simeq 10^8$, on the other hand one can argue that only the electrons close to Fermi surface should matter what gives $n_0\simeq 10^4$. The main assumption behind the simple Hamiltonian (\ref{cpbham}) is that all Cooper pairs occupy a single quantum state, similarly to the BEC picture and therefore only the number of Cooper pairs matters.
\par
Under the assumption $n_0 >> 1$ and restricting to the states with $|n-n_0| << n_0$ the Hamiltonian (\ref{cpbham}) can be rewritten in terms of  spin variables $\hat{J}_k , k=1,2,3$
satisfying standard relations
\begin{equation}
[\hat{J}_k , \hat{J}_l] = i\sum_{m=1}^3\epsilon_{klm}\hat{J}_m\ ,\ \hat{J}_1^2 +\hat{J}_2^2 + \hat{J}_3^2 = j(j+1)
\label{spin}
\end{equation}
with $j=n_0 $. The Josephson Hamiltonian  reads now
\begin{equation}
\hat{H} = 4E_C( \hat{J}_3 - n_g)^2  - \frac{E_J}{2j}\hat{J}_1 .
\label{cpbham1}
\end{equation}
An essentially equivalent model Hamiltonian has a form of the nonlinear oscillator's one
\begin{equation}
\hat{H} = 4E_C \bigl( \hat{a}^{\dagger} \hat{a} -(n_0 + n_g)\bigr)^2  - \frac{E_J}{2\sqrt{n_0}}(\hat{a} +\hat{a}^{\dagger})
\label{osc}
\end{equation}
where $[\hat{a} ,\hat{a}^{\dagger}]=1$. In the following only the large spin model is used.

\par
The device is controlled by external electromagnetic fields which  are coupled to the net electric charge $\hat{Q}= 2e \hat{J}_3 $ and to the electric current $\frac{d\hat{Q}}{dt}= i[\hat{H},\hat{Q}]\sim \hat{J}_2$. Hence the control Hamiltonian is given by 
\begin{equation}
\hat{H}_c(t) = h_2(t)\hat{J}_2 + h_3(t)\hat{J}_3 .
\label{Rycon1}
\end{equation}

To describe the influence of an environment one should notice that the leading contribution to the system-bath interaction is always of the form similar to (\ref{Rycon1}) with the external fields $h_k(t)$ replaced by bath's operators $\hat{B}_k ; k= 1,2,3$ and taking into account that the fluctuations of the tunneling rate produce the term with $k=1$.
\par
Under quite general conditions the reduced dynamics of the spin in Markovian approximation \cite{Ali} is given by the master equation  for the reduced density matrix $\hat{\rho}$
\begin{equation}
\frac{d}{dt}\hat{\rho} = -i[\hat{H} , \hat{\rho} ] + \sum_{k,l=1}^3 c_{kl}\bigl([\hat{J}_k \hat{\rho},\hat{J}_l] + [\hat{J}_k , \hat{\rho} \hat{J}_l]\bigr)
\label{ME}
\end{equation}
where $\hat{H}$ is given by (\ref{cpbham1}) and the positively defined matrix $[c_{kl}]$ depends on the details of the reservoir.
The stability of the initial pure state $|\psi\rangle$ can be characterized by the initial decay of purity
\begin{equation}
\Gamma(\psi)\equiv -\frac{d}{dt}\mathrm{Tr}\hat{\rho}^2|_{t=0} = 2\sum_{k,l=1}^3 c_{kl}\bigl(\langle\psi, \hat{J}_k \hat{J}_l \psi\rangle -\langle\psi \hat{J}_k\psi \rangle\langle\psi \hat{J}_l\psi \rangle\bigr)
\label{pure}
\end{equation}
The RHS of (\ref{pure}) can be treated as a dispersion of the total spin modified by some weights $\gamma_k$- eigenvalues
of the relaxation matrix $[c_{kl}]$. Therefore, one can see that the purity of the eigenstates $|j,m\rangle$ decays with
the rate $\Gamma (|j,m\rangle)\simeq \gamma j^2$ . By  semiclassical states we mean the states which minimize the dispersion of the total spin to the value of the order $\sim j$. The decay rate of such states is of the order of $\gamma j$ ($\gamma$ is a typical value of the relaxation constants $\gamma_k$). It follows that for large enough spins only semiclassical states are relatively stable and can be observed \cite{Ben}.

This type of "dequantization" is supposed to be the main mechanism of emergence of classical world and provides the solution for the problem of molecular structure, existence of deformed nuclei and generally the absence of \emph{Schr\"odinger cat states}. A nice illustration of this mechanism is provided by the experiments with atomic BEC placed in a double-well
potential \cite{Es}. The Hamiltonian of such a system can be approximated by the Josephson one (\ref{cpbham1}) where the observable $\hat{J}_3$ corresponds to the excess number of atoms with respect to an equilibrium value $n_0$ in a chosen well. The semiclassical fluctuations of the number of bosons in a given well $\sim\sqrt{n_0}$ are clearly observed.
\par
On the other hand, for CPBs such large fluctuations of the number of Cooper pairs are not observed, on the contrary, superpositions of states which differ by a single Cooper pair are visible in numerous experiments. This suggests that perhaps the model based on the Josephson Hamiltonian (\ref{cpbham1}) is not adequate.

\section{Lattice gas model}

In this model one assumes that due to strong Coulomb repulsion between Cooper pairs on the small electrode  they cannot occupy a single quantum state but rather a set of localized sites with the occupation number 1 or 0.
The tunneling to a large electrode is possible only from the sites placed at the boundary of the small electrode which is close to the large one.  The number of these relevant sites, denoted by $m_0$ should be roughly proportional to the square root of the total number of Cooper pairs $n_0$. Taking $n_0 \simeq 10^8 - 10^4$ one obtains $m_0\simeq 10^4-10^2$. Introducing the occupation number operators $\hat{n}_k = (\hat{\sigma}^z_k +1)/2 ; k=1,2,...,m_0$ with eigenvalues $0,1$ and spin-1/2 Pauli matrices
$\hat{\sigma}^{\alpha}_k, \alpha = x,y,z ,+, -, k= 1,2,...,m_0$ 
one can propose the following model Hamiltonian for this system
\begin{equation}
\hat{H} = 4E_C \bigl(\sum_{k=1}^{m_0}\hat{n}_k - (m_0 + n_g)\bigr)^2  - \frac{E_J}{2}\sum_{k=1}^{m_0}(\xi_k \hat{\sigma}^-_k + \bar{\xi}_k \hat{\sigma}^+_k ).
\label{latham}
\end{equation}
Here the first term describes Coulomb repulsion and $n_g\in [0, 1)$ is a fine tuning control parameter. The second term describes tunneling from and to  individual sites
with the normalization of local tunneling parameters $\sum |\xi_k|^2 =1$. Using the standard assumption $E_C > E_J >> kT$  one can restrict the effective Hilbert space to the lowest energy spectrum sector $\mathcal{H}_{\mathrm{eff}}$ spanned by the following $m_0+1$ basis eigenvectors of the operators $\hat{n}_k$
\begin{equation}
|0\rangle \equiv |1,1...,1\rangle\  ,\  |k\rangle \equiv |\underbrace{1,...,1,0}_k,1,...,1\rangle\ ,\ k=1,2,...,m_0. 
\label{basis}
\end{equation}
Introducing a normalized  vector and two projectors in $\mathcal{H}_{\mathrm{eff}}$
\begin{equation}
|\xi\rangle\equiv \sum_{k=1}^{m_0}\xi_k|k\rangle \ ,\ 
\hat{P}=|0\rangle\langle 0|+|\xi\rangle\langle\xi|\ ,\  \hat{P}^{\bot} = \hat{I}-\hat{P}
\label{proj}
\end{equation}
one can write down the Hamiltonian (\ref{latham}) restricted to $\mathcal{H}_{\mathrm{eff}}$ and up to an irrelevant constant as
\begin{equation}
\hat{H} = \frac{E(n_g)}{2}(|\xi\rangle\langle \xi| - |0\rangle\langle 0| -\hat{P}) - \frac{E_J}{2}(|\xi\rangle\langle 0| + |0\rangle\langle\xi|)\ ,\ E(n_g)= 4E_C(1-2n_g) .
\label{hamef}
\end{equation}
The energy levels of (\ref{hamef}) consist of the $m_0-1$ manifold of degenerated levels of the energy equal to zero given by the projector $\hat{P}^{\bot}$ and two levels $|\pm\rangle$ of positive and negative energies $E_{\pm}$
\begin{equation}
|+\rangle = \cos\frac{\theta}{2}|\xi\rangle -\sin\frac{\theta}{2}|0\rangle \ ,\ |-\rangle = \cos\frac{\theta}{2}|0\rangle +\sin\frac{\theta}{2}|\xi\rangle \ ,\ E_{\pm} = \pm\frac{1}{2}\sqrt{E(n_g)^2 + E_J^2} - \frac{1}{2}E(n_g)
\label{eigenvec}
\end{equation}
where $\theta$ is defined by $\cos\theta=E(n_g)/\sqrt{E(n_g)^2 + E_J^2}$.

The external control is performed by the coupling through the total electric charge operator $\hat{Q}$ and the total electric current $\hat{J}$ which, when restricted to $\mathcal{H}_{\mathrm{eff}}$ reads
\begin{equation}
\hat{Q} = e(|\xi\rangle\langle\xi|-|0\rangle\langle 0|  + \hat{P}^{\bot})\ ,\ \hat{J}= i[\hat{H},\hat{Q}] = ieE_J( |0\rangle\langle\xi|-|\xi\rangle\langle 0|).
\label{echarge}
\end{equation}
Obviously, if the system is completely isolated the \emph{qubit space} spanned by $|0\rangle $ and $|\xi\rangle $ is invariant with respect to the Hamiltonian and the external control yielding the usual model of  charge
qubit. We denote its qubit observables by
\begin{equation}
\hat{\sigma}_{\xi}^+=|+\rangle\langle -| \ ,\hat{\sigma}_{\xi}^-= (\hat{\sigma}_{\xi}^+)^{\dagger}\ ,\ \hat{\sigma}_{\xi}^z = (|+\rangle\langle +| - |-\rangle\langle -|).
\label{qubit}
\end{equation}
\par
In contrast to the large spin model the coupling of the lattice gas model to a bath has individual and local character. The sufficiently general form of the system-bath interaction Hamiltonian is given by
\begin{equation}
\hat{H}_{\mathrm{int}} = \sum_{k=1}^{m_0}\sum_{\alpha=x,y,z} \hat{\sigma}_k^{\alpha}\otimes \hat{B}_k^{\alpha}  
\label{hamint}
\end{equation}
restricted to the subspace $\mathcal{H}_{\mathrm{eff}}\otimes\mathcal{H}_{\mathrm{bath}} $. One can assume that the bath operators $\hat{B}_k^{\alpha}$ correspond to different independent constituents of the bath ("private baths" picture). Applying now the standard weak coupling limit technique and assuming that the temperature of the bath is zero
and the private baths are identical one obtains the following master equation for the reduced density matrix of the system \cite{Ali}
\begin{align}
\frac{d}{dt}\hat{\rho} = &-i[\hat{H} , \hat{\rho} ] + \lambda([\hat{\sigma}^-_{\xi} \hat{\rho},\hat{\sigma}^+_{\xi} ] + [\hat{\sigma}^-_{\xi} , \hat{\rho} \hat{\sigma}^+_{\xi} ]\bigr)
-\nu[\hat{\sigma}^z_{\xi},[\hat{\sigma}^z_{\xi},\hat{\rho}]]\\
 &+ \mu_1\sum_{k=1}^{m_0-1}([\hat{a}_k \hat{\rho},\hat{a}_k^{\dagger} ] + [\hat{a}_k, \hat{\rho}\hat{a}_k^{\dagger} ])
 + \mu_2\sum_{k=1}^{m_0-1}([\hat{b}_k \hat{\rho},\hat{b}_k^{\dagger} ] + [\hat{b}_k, \hat{\rho}\hat{b}_k^{\dagger} ]).
\label{ME1}
\end{align}
Here
\begin{equation}
\hat{a}_k = \hat{P}^{\bot}|k\rangle\langle +|\ ,\   \hat{b}_k =|-\rangle \langle k|\hat{P}^{\bot} 
\label{Dav}
\end{equation}
and the decay rates $\lambda, \mu_1 , \mu_2\geq 0$ and "pure dephasing" rate $\nu$  are given by standard expressions involving Fourier transforms of the autocorrelation functions for baths operators taken at frequencies corresponding to
proper energy differences.
\par
One can obtain from (\ref{ME1}) the Pauli Master Equation for  the level occupation probabilities $p_{\pm} = \langle\pm|\hat{\rho}|\pm\rangle $ , $p_0 = \mathrm{Tr}(\hat{\rho}\hat{P}^{\bot}_{\xi})$:
\begin{align}
\frac{dp_+}{dt} &= -2[\lambda +(m_0 -1)\mu_1] p_+ \\
\frac{dp_0}{dt} &= 2(m_0 -1)\mu_1 p_+ -2\mu_2 p_0 \\
\frac{dp_-}{dt} &= 2\lambda p_+ + 2\mu_2 p_0
\label{pau}
\end{align}
and the closed equation for the qubit coherence $z = \mathrm{Tr}(\hat{\rho}\hat{\sigma}^+_{\xi})$
\begin{equation}
\frac{dz}{dt} = (i\omega -[\lambda + 4\nu +(m_0 -1)\mu_1])z\ ,\ \omega =\sqrt{E(n_g)^2 + E_J^2}
\label{coh}
\end{equation}
\par
One can notice that the leading damping effects in (\ref{pau}, \ref{coh}) are proportional to $\Gamma = (m_0-1)\mu_1>> \mu_1,\mu_2 ,\nu, \lambda$. Taking only them into account one obtains an approximative solution on the time scale of the order
$\Gamma^{-1}$
\begin{equation}
p_+(t) =p_+(0)e^{-\Gamma t}\ ,\ p_-(t)= p_-(0) \ ,\ p_0(t) = p_0(0)+ p_+(0)[1-e^{-\Gamma t}]\ ,\ z(t) = z(0)\exp\{(i\omega -\Gamma/2)t\}.
\label{sol}
\end{equation}
In this approximation the third, highly degenerated level acts a  "probability sink" for the qubit.  The long time behavior is given by
\begin{equation}
 t>>\Gamma^{-1}\ ,\ p_+(t) =0\ ,\  p_0(t) = p_+(0)e^{-2\mu_2 t}\ ,\ p_-(t) = 1- p_0(t)\ ,\ \ z(t) = 0 .
\label{sol1}
\end{equation}
The only directly measured observable is the net charge 
\begin{equation}
Q(t)=\mathrm{Tr}(\hat{\rho}(t)\hat{Q})= ( z(t) + \bar{z}(t))\sin\theta - (1-p_0(t))\cos\theta + p_0(t). 
\label{charge}
\end{equation}
We are interested also in the time dependence of the averaged energy given by
\begin{equation}
{\cal E}(t)=\mathrm{Tr}(\hat{\rho}(t)\hat{H})= E_+ p_+(t) + E_-p_-(t). 
\label{energy}
\end{equation}
Assume, for simplicity, that one prepares a system in a pure state $|\xi\rangle$ and choose $n_g = 1/2$ ($\sin\theta=1$) what corresponds to the experimental setting of \cite{Nak}, \cite{Bla}. Combining the formulas (\ref{charge}) and (\ref{sol1}) one obtains
the evolution of the mean charge for short times
\begin{equation}
 Q(t)=  \Bigl[\bigl(\cos(\omega t) +\frac{1}{2} e^{-\frac{\Gamma t}{2}} \Bigr]e^{-\frac{\Gamma t}{2}} + \frac{1}{2}.
\label{charge1}
\end{equation}
Although the predicted form of "coherent charge oscillations"  differs from that obtained for the 2-level model 
\begin{equation}
Q(t)= (2 \cos\omega t) e^{-\frac{t}{T_{2}}} ,
\label{charge2}
\end{equation}
with the \emph{decoherence time} $T_2$, both curves can be fitted to the experimental data. Moreover, both curves give
the same Lorentzian frequency spectrum around $\omega$ when $T_2 = 2/\Gamma$.
\par 
The measurements of energy relaxation are usually performed for $n_g \simeq 0$ ($\sin\theta\simeq 0$). As $E_C > E_J$   hence
\begin{equation}
E_+\simeq 0\ ,\ E_-\simeq -4E_C\ . 
\label{energy1}
\end{equation}
It follows from (\ref{sol1}),(\ref{energy}) and (\ref{energy1}) that the energy decays to its lowest value on the long time scale
\begin{equation}
{\cal E}(t)= 4E_c \bigl(p_+(0)e^{-2\mu_2 t} -1\bigr). 
\label{energy2}
\end{equation}
The relation (\ref{energy2}) can explain observed in \cite{Leh} long relaxation times corresponding to $T_1 = (2\mu_2)^{-1}$ in contrast to a much faster decay of $p_+(t)$ with  decay time ${T_1}'= \Gamma^{-1}$ which seems to determine the results of \cite{Bla}. The ratio $T_1 / T_2 \simeq m_0/2$ estimated from the experimental data of \cite{Leh} gives a reasonable value $m_0 \simeq 10^3$.

\section{Conclusions}
A new model of a Cooper pair box is proposed, based on the picture of a large number $m_0$ of localized states ("lattice sites") which are occupied by at most one Cooper pair for each site and from which the Cooper pairs can tunnel to a large superconducting reservoir and back. This model is compared with the standard model of a large spin $j= n_0$ system, where $n_0$ is an average number of Cooper pairs which occupy a single quantum state delocalized over the whole superconducting island. The main difference between both models is due to the interaction with an environment.
For the large spin model all Cooper pairs interact collectively with the bath what makes the states with a fixed charge very unstable with decoherence rates $\sim n_0^2$. The relatively stable states
with life-times $\sim n_0$ possess semiclassical character and yield large charge fluctuations 
of the order $\sqrt{n_0}$. This behavior is confirmed for the atomic Bose-Einstein condensate in a double well, which is described by the similar mathematical model. It is very difficult to explain why this mechanism should not work for a CPB and why the well defined charge states remains relatively stable. 
\par
The lattice gas model suggests a localized coupling of any site to its individual heat bath what at the low temperature regime produces charge fluctuations of the order ${\cal O}(1)$ in agreement with experiments. However, the many-body character of the system is still present.
The relevant part of the Hilbert space is spanned not only by a pair of Hamiltonian eigenstates
corresponding to a qubit but also by $m_0 -1$ degenerated states which form a probability sink for the excited state of the qubit with the decay rate $\sim (m_0 -1)$.  The predicted form of the damped charge oscillations differs from the standard formula providing a possible experimental test of the model. This model explains also a huge discrepancy between different data on relaxation times, introducing natural two time scales for dissipative processes in the system.
\par
Finally, one should mention another difference between the standard qubit model of  CPB
and the lattice gas model. In the former the effective 2-dimensional Hilbert space is spanned by the vectors 
corresponding to nondegenerated eigenstates of a charge operator. In the later model, one of the qubit states is a superposition of a large number of degenerated charge eigenstates with the probability amplitudes determined by  local tunneling rates which can vary in time. Therefore, one can expect to observe a slow random drift of that  state over the $m_0$ dimensional Hilbert subspace. This means that not only the probability leaks from the qubit but the qubit itself does not correspond to a fixed well-defined 2-dimensional subspace.


\end{document}